\documentclass[twocolumn,superscriptaddress,aps,prb]{revtex4}
\usepackage{graphicx,amsmath,amssymb}
\usepackage{epstopdf}
\usepackage{amssymb}
\usepackage{grffile}
\usepackage[usenames]{color}
\usepackage{indentfirst}
\usepackage{float}
\usepackage{color}
\usepackage{mathrsfs}
\usepackage{hyperref}
\usepackage{dcolumn}
\usepackage{bm}

\begin{document}

\title{Analytically solvable model to the spin Hall effect with Rashba and
Dresselhaus spin-orbit couplings}
\author{Rui Zhang}
\affiliation{Department of Physics, Chongqing University, Chongqing
401331, People's Republic of China}
\author{Yuan-Chuan Biao}
\affiliation{Department of Physics, Chongqing University, Chongqing
401331, People's Republic of China}
\author{Wen-Long You}
\affiliation{College of Science, Nanjing University of Aeronautics and Astronautics,
Nanjing 211106, People's Republic of China}
\author{Xiao-Guang Wang}
\affiliation{Department of Physics, Zhejiang University, Hangzhou 310027, People's
Republic of China}
\author{Yu-Yu Zhang}
\email{yuyuzh@cqu.edu.cn}
\affiliation{Department of Physics, Chongqing University, Chongqing
401331, People's Republic of China}
\author{Zi-Xiang Hu}
\email{zxhu@cqu.edu.cn}
\affiliation{Department of Physics, Chongqing University, Chongqing
401331, People's Republic of China}

\begin{abstract}
When the Rashba and Dresslhaus spin-orbit coupling are both presented for a two-dimensional
electron in a perpendicular magnetic field, a striking resemblance to
anisotropic quantum Rabi model in quantum optics is found. We perform a
generalized Rashba coupling approximation to obtain a solvable Hamiltonian 
by keeping the nearest-mixing terms of Laudau
states, which is reformulated in the similar form to that
with only Rashba coupling. Each Landau state becomes a new displaced-Fock state
with a displacement shift instead of the original Harmonic oscillator Fock
state, yielding eigenstates in closed form. Analytical energies are
consistent with numerical ones in a wide range of coupling strength even for
a strong Zeeman splitting. In the presence of an electric field,
the spin conductance and the charge conductance obtained analytically are
in good agreements with the numerical results. As the component of the Dresselhaus coupling
increases, we find that the spin Hall conductance exhibits a
pronounced resonant peak at a larger value of the inverse of the magnetic
field. Meanwhile, the charge conductance exhibits a series of
plateaus as well as a jump at the resonant magnetic field. Our method provides an
easy-to-implement analytical treatment to two-dimensional electron gas
systems with both types of spin-orbit couplings.
\end{abstract}

\date{\today }
\maketitle

\section{Introduction}

Spin-orbit coupling (SOC) enables a wide variety of fascinating phenomena, 
which brings out a new growing research field of spin-orbitronics, a branch
of spintronics that focuses on the manipulation of the electron spin degree
of freedom~\cite{Fabian, Chappert}. A prominent example is the spin Hall
effect, which is the conversion of a spin unpolarized charge current into a
net spin current without charge flow~\cite{Dyakonov,Hirsch}. It has been
discussed intensively~\cite{zhang2003,niu2004,
NiuRMP,Kato,Wunder,Wunderlich,Valenzuela,JairoRMP}, in which an electric
field induces a transverse spin current. In the presence of a perpendicular
magnetic field, the interplay of Zeeman coupling and various spin-orbit
interactions has stimulated a lot of discussions on resonance spin Hall
conductance~\cite{shen2004,tagliacozzo2008,engel07}, which may has potential
applications in spintronics.

There are basically two types of SOCs in nature, i.e., the Rashba term with
structural inversion asymmetry~\cite{rashba1960} and the Dresselhaus term
due to the bulk inversion asymmetry~\cite{dresselhaus1955} or interface
inversion asymmetry~\cite{IIA}. Usually, both types of SOC coexist in a
material, such as GaAs-AlGaAs quantum wells and heterostructures~\cite%
{jusserand,kim2016,ishizaka2011,Manchon2015,lommer1988,tretyak2001}, but
which one plays a major part depends on the properties of the material. It
has been recognized that Rashba and Dresselhaus SOC can interfere with each
other, and leads to a number of interesting phenomenon by tuning the ratio
between them, such as anisotropic transport~%
\cite{loss2003}, spin splitting~\cite{ganichev04,averkiev1999}, control of
spin precession~\cite{park2020} and light scattering~\cite{gelfert2020}. It
has been applied experimentally by various ways, such as external electric
field, changing temperature, or inserting extra layer etc~\cite%
{Ohno,Karimov,Dhrmann,Belkov,Muller,Balocchi,Hern}.

In previous work, many efforts have been devoted to the spin Hall effect
with the Rashba SOC in two-dimensional electron gas systems ($2$DEGs), which
is caused by the Laudau level crossing near the Fermi energy. A zero-field
spin splitting induced by the Rashba SOC competes with the Zeeman splitting
in presence of a magnetic field, and then compromise a resonant spin Hall
effect at certain magnetic field~\cite{nitta1997,heida1998}. In contrast,
Dresselhaus SOC enhances the Zeeman splitting and results in a suppression
to the resonance~\cite{shen05}. There is an analytical solution for the
system with only either Rashba or Dresselhaus coupling~\cite{shen05,chang06}%
. While considering both of them together, an analytical solution is
currently not available due to the absence of a full closed-form solution.
The perturbation method has been adopted to give rise to an approximated
results, which is valid a small ratio of the Zeeman energy to the cyclotron
frequency~\cite{shen05,sun2013}. To pursue the intrinsic spin Hall effect
induced by the coexistence of the Dresselhaus and Rashba SOCs, it is
desirable to develop theories which can solve systems including both types
of the SOCs.

In this work we develop a generalized Rashba SOC approximation (GRSOCA) to give 
an analytical solution to the $2$DEGs with both the
Rashba and Dresselhaus SOCs in a magnetic field. Using a displacement transformation and an expansion
of even and odd functions of Laudau states up to the nearest-mixing terms, a
reformulated Hamiltonian of the same form as that with only Rashba term is
obtained, resulting in eigenstates in closed form in the transformed
displaced-Fock subspace. The novel displaced-Fock state for each Landau
level involves the mixing of infinite Laudau level states induced by two
types of SOCs, exhibiting an improvement over original Fock states. Energy
levels are obtained explicitly for arbitrary strengths of both types of
SOCs, which agree well with numerical results in a wide range of coupling
strength even for a strong Zeeman energy. By comparing to the system with
only Rashba SOC, we find that the spin Hall conductance exhibits a
pronounced resonant peak at a larger value of the inverse of the magnetic
field, which arises from the contributions of the Dresselhaus SOC. Resonance
originates from the energy degeneracy near Fermi energy, where the
eigenstates consist of the $n$th displaced-Fock state of spin up and $n+1$th
displaced-Fock state of spin down. A series of plateaus of the charge conductance are observed, 
and a jump occurs at the resonant point, which fit well with the numerical
ones.

The paper is outlined as follows. In Sec.~II, we derive expressions for the
quantized Hamiltonian of the $2$DEGs with the Rashba and Dresselhaus SOC in
a magnetic field. In Sec.~III, we obtain the analytical solution of the
effective Hamiltonian for arbitrary ratio between the Rashba and Dresselhaus
SOCs. In Sec.~IV, we study the charge and spin Hall conductances
analytically with the first-order corrections when an electric field is
applied. Finally, a brief summary is given in Sec.~V.

\section{Hamiltonian}

We consider a single electron in a two-dimensional system subjected to a
perpendicular magnetic field $\vec{B}=-B\hat{e}_{z}=\nabla \times \vec{A}$,
which is confined in the $x-y$ plane of an area $L_x\times L_y$. The Hamiltonian
in the presence of spin-orbit coupling is given by ($\hbar =1$)
\begin{equation}  \label{originH0}
H_{0}=\frac{1}{2m}(\vec{p}+\frac{e}{c}\vec{A})^{2}-\frac{1}{2}g\mu
_{B}B\sigma _{z}+H_{so},
\end{equation}
where $g$ is the Lande factor of the electron with the effective mass $m$, $%
\mu _{B}$ is the Bohr magneton, and $\sigma_{k}$ are the Pauli matrices. The
Laudau gauge is chosen as $\vec{A}=B\vec{r}\times \hat{e} _{z}=(yB,0,0)$.
The spin-orbit Hamiltonian includes the Rashba SOC and the linear
Dresselhaus SOC, $H_{so}=H_{D}+H_{R}$ with $H_{R}=\alpha(\Pi _{x}\sigma _{y}-\Pi _{y}\sigma
_{x})$ and $H_{D}=\beta(\Pi _{x}\sigma _{x}-\Pi
_{y}\sigma _{y})$, where the canonical momentum is $\vec{\Pi} =\vec{p}+e\vec{%
A}/c$. The Rashba SOC $H_{R}$ originates from the structure inversion
asymmetry of the semiconductor material and the coupling strength $\alpha $
can be tuned by an electric field. While the coefficient $\beta $ of the
Dresselhaus term $H_{D}$ is determined by the geometry of the
hetereo-structure that stems from the bulk-inversion asymmetry of the
semiconductor material.

Due to the gauge choice, the system is translationally invariant along the $%
x $ direction, and $p_{x}=k$ is a good quantum number. The orbit part of
wave function is obtained as $\psi (x,y)=\exp (ikx)\varphi (y-y_{0})$, where
$\varphi (y-y_{0})$ is the harmonic oscillator wave function with the orbit
center coordinate $y_{0}=l_{b}^{2}k$ and the magnetic length $l_{b}=\sqrt{%
\hbar c/eB}$. By introducing the ladder operator $a=(\Pi
_{x}+i\Pi _{y})l_{b}/\sqrt{2}$ for the harmonic oscillator
\begin{equation}
a=\frac{1}{\sqrt{2}l_{b}}[y+\frac{c(p_{x}+ip_{y})}{eB}],
\end{equation}%
one obtains the Hamiltonian
\begin{equation}  \label{origha0}
H_{0}=H_{\mathtt{RSOC}}+\frac{\sqrt{2}\beta }{l_{b}}(a^{\dagger }\sigma
_{-}+a\sigma _{+}),
\end{equation}
\begin{equation}
H_{\mathtt{RSOC}}=\omega (a^{\dagger }a+\frac{1}{2})-\frac{\Delta }{2}%
\sigma _{z}+i\frac{\sqrt{2}\alpha }{l_{b}}(a\sigma _{-}-a^{\dagger }\sigma
_{+}),
\end{equation}
where $\omega =eB/mc$ is the cyclotron frequency and $\Delta =g\mu _{B}B$ is
the Zeeman splitting. When only the Rashba SOC term is present, i.e. $\beta=0
$, the Hamiltonian is reduced to $H_{\mathtt{RSOC}}$ which can be written in
a matrix form in the basis \{$|\downarrow ,n\rangle $, $|\uparrow
,n+1\rangle \}$
\begin{equation}
H_{\mathtt{RSOC}}=\left(
\begin{array}{cc}
\omega(n+1/2)+\Delta /2 & i\sqrt{2(n+1)}\alpha /l_{b} \\
-i\sqrt{2(n+1)}\alpha /l_{b} & \omega(n+1+1/2)-\Delta /2%
\end{array}%
\right) .  \label{rsoc}
\end{equation}%
This Hamiltonian, like a Rabi model with only the counter-rotating wave term
in the quantum optics, can be solved analytically in closed subspace,
which is so-called Rashba SOC (RSOC) approximation.
However, once including
the additional Dresselhaus SOC term (rotating wave term in Rabi model), the
subspace related to $n$ is not closed, rendering the complication of the
solution. In that case, each Landau level is coupled to infinite number of
other Landau levels, and thus the exact analytic solution is not available.

\section{Analytical solution}

Following previous section, the Hamiltonian (~\ref{origha0}) of a two-dimensional
electron with both the Rashba and Dresselhaus couplings can map onto the
anisotropic Rabi model in quantum optics, which have been studied
extensively by various approximate analytical solutions~\cite%
{chen12,irish07,zhang16}. The crucial is to establish a new set of basis states.
  
To facilitate the study, we write the Hamiltonian
as
\begin{eqnarray}\label{originalH}
H_{0} &=&\omega (a^{\dagger }a+\frac{1}{2})-\frac{\Delta }{2}\sigma
_{z}+g_{1}\sigma _{x}(a^{\dagger }e^{-i\theta }+ae^{i\theta })  \notag\\
&&-g_{1}i\sigma _{y}(a^{\dagger }e^{i\theta }-ae^{-i\theta }),
\end{eqnarray}%
with $g_{1}=\sqrt{2}\sqrt{\beta^{2}+\alpha ^{2}}/l_{b}$, and $e^{i\theta
}=(\beta +i\alpha )/\sqrt{\beta ^{2}+\alpha ^{2}}$. By performing a unitary
transformation $U=\exp [\sigma _{x}(a^{\dagger }\gamma -a\gamma ^{\ast })]$
with a dimensionless variational displacement $\gamma $ ($\gamma ^{\ast }$),
we obtain a transformed Hamiltonian $H_{1}=UHU^{\dagger
}=H_{0}^{\prime }+H_{1}^{\prime }$ ,
\begin{eqnarray}\label{transh0}
H_{0}^{\prime } &=&\omega a^{\dagger }a+\eta _{0}+\sigma _{z}\{\eta
_{1}\cosh [2(a^{\dagger }\gamma -a\gamma ^{\ast })]  \notag \\
&&+g_{1}\sinh [2(a^{\dagger }\gamma -a\gamma ^{\ast })](a^{\dagger
}e^{i\theta }-ae^{-i\theta })\},
\end{eqnarray}%
\begin{eqnarray}\label{transh1}
H_{1}^{\prime } &=&\sigma _{x}[a^{\dagger }(g_{1}e^{-i\theta }-\omega \gamma
)+a(g_{1}e^{i\theta }-\omega \gamma ^{\ast })]  \notag \\
&&+i\sigma _{y}\{-\eta _{1}\sinh [2(a^{\dagger }\gamma -a\gamma ^{\ast
})]-g_{1}\cosh [2(a^{\dagger }\gamma  \notag \\
&&-a\gamma ^{\ast })](a^{\dagger }e^{i\theta }-ae^{-i\theta })\},
\end{eqnarray}%
where $\eta _{0}=\omega /2-g_{1}(\gamma ^{\ast }e^{-i\theta }+\gamma
e^{i\theta })+\omega \gamma \gamma ^{\ast }$ and $\eta _{1}=-\Delta
/2-g_{1}(\gamma e^{-i\theta }-\gamma ^{\ast }e^{i\theta })$ in Appendix A. The
displacement shift $\gamma $ ($\gamma ^{\ast }$) is associated with the
Rashba SOC and Dresselhaus SOC strengths, which captures the displacement of
the harmonic oscillator states for essential physics.

Since the even hyperbolic cosine function can be expanded as $\cosh
[2(a^{\dagger }\gamma -a\gamma ^{\ast })]=1+\frac{1}{2!}[2(a^{\dagger
}\gamma -a\gamma ^{\ast })]^{2}+\frac{1}{4!}[2(a^{\dagger }\gamma -a\gamma
^{\ast })]^{4}+\cdots $, it is approximated by keeping the terms which only
contain the number operator $\hat{n}=a^{\dagger }a$ as~\cite{zhang16}
\begin{equation}
\cosh [2(a^{\dagger }\gamma -a\gamma ^{\ast })]=G(a^{\dagger }a)+O(\gamma
^{2}\gamma ^{\ast 2}).
\end{equation}%
The coefficient $G(a^{\dagger }a)$ can be expressed in the harmonic
oscillator basis $|n\rangle $ as
\begin{equation}
G_{n,n}=\left\langle n\right\vert \cosh [2(a^{\dagger }\gamma -a\gamma
^{\ast })]\left\vert n\right\rangle =e^{-2\gamma \gamma ^{\ast
}}L_{n}(4\gamma \gamma ^{\ast }),
\end{equation}%
with the Laguerre polynomials $L_{n}^{m-n}(x)=\sum_{i=0}^{\min
\{m,n\}}(-1)^{n-i}\frac{m!x^{n-i}}{(m-i)!(n-i)!i!}$. Here, the higher order
excitations such as $a^{\dagger 2}$, $a^{2}$, $\cdots$ are neglected in the
approximation. Similarly, we expand the odd function $\sinh [2(a^{\dagger
}\gamma -a\gamma ^{\ast })]$ by keeping the one-excitation terms as
\begin{equation}
\sinh [2(a^{\dagger }\gamma -a\gamma ^{\ast })]=R(a^{\dagger }a)a^{\dagger
}-aR(a^{\dagger }a)+O(\gamma ^{3}\gamma ^{\ast 3}).
\end{equation}%
Since the terms $R(a^{\dagger }a)a^{\dagger }$ and $aR(a^{\dagger }a)$ are
conjugated to each other, which corresponds to create and eliminate a single
excitation of the oscillator, we define
\begin{eqnarray}
R_{n,n+1} &=&-\frac{1}{\sqrt{n+1}}\left\langle n\right\vert \sinh
[2(a^{\dagger }\gamma -a\gamma ^{\ast })]\left\vert n+1\right\rangle  \notag
\\
&=&\frac{2\gamma ^{\ast }}{n+1}e^{-2\gamma \gamma ^{\ast }}L_{n}^{1}(4\gamma
\gamma ^{\ast }) =R_{n+1,n}^{\ast}.
\end{eqnarray}
Similarly, the other operators can be expanded by keeping leading terms as
follows:
\begin{equation}
\sinh [2(a^{\dagger }\gamma -a\gamma ^{\ast })](a^{\dagger }e^{i\theta
}-ae^{-i\theta })=F(a^{\dagger }a)+O(\gamma ^{2}\gamma ^{\ast 2}),
\end{equation}%
\begin{eqnarray}
\cosh [2(a^{\dagger }\gamma -a\gamma ^{\ast })](a^{\dagger }e^{i\theta
}-ae^{-i\theta }) \approx T(a^{\dagger }a)a^{\dagger }-aT(a^{\dagger }a),
\notag \\
\end{eqnarray}
where the coefficients can be expressed in terms of the oscillator basis $%
|n\rangle $
\begin{eqnarray}
F_{n,n} &=&\left\langle n\right\vert \sinh [2(a^{\dagger }\gamma -a\gamma
^{\ast })](a^{\dagger }e^{i\theta }-ae^{-i\theta })\left\vert n\right\rangle
\notag \\
&=&-e^{i\theta }(n+1)R_{n,n+1}-ne^{-i\theta }R_{n,n-1},
\end{eqnarray}%
\begin{eqnarray}
T_{n,n+1} &=&\frac{-\left\langle n\right\vert \cosh [2(a^{\dagger }\gamma
-a\gamma ^{\ast })](a^{\dagger }e^{i\theta }-ae^{-i\theta })\left\vert
n+1\right\rangle}{\sqrt{n+1}}  \notag \\
&=&e^{-i\theta }G_{n,n}-\frac{\sqrt{n+2}}{\sqrt{n+1}}e^{i\theta }G_{n,n+2},
\end{eqnarray}%
with $G_{n,n+2}=\left\langle n\right\vert \cosh [2(a^{\dagger }\gamma
-a\gamma ^{\ast })]\left\vert n+2\right\rangle =(2\gamma ^{\ast })^{2}\exp
[-2\gamma \gamma ^{\ast }]L_{n}^{2}(4\gamma \gamma ^{\ast })/\sqrt{(n+1)(n+2)%
}$.

Finally, we obtain the reformulated Hamiltonian $\tilde{H}_{1}=\tilde{H}_{0}+%
\tilde{H}_{D}$, consisting of
\begin{eqnarray}  \label{transh0}
\tilde{H}_{0}=\omega a^{\dagger }a+\eta _{0}+\sigma _{z}\widetilde{\Delta }+\widetilde{\alpha}a^{\dag }\sigma _{+}+\widetilde{\alpha}^{*}a\sigma _{-},
\end{eqnarray}
\begin{equation}
\tilde{H}_{D}=\widetilde{\beta}a^{\dag }\sigma _{-}+\widetilde{\beta}^{*}a\sigma _{+},
\end{equation}
where the Zeeman energy is renormalized as  $\widetilde{\Delta }=\eta
_{1}G(a^{\dagger }a)+g_{1}F(a^{\dagger }a)$, the effective Rashba and Dresselhaus SOCs strength are derived as $\widetilde{\alpha}=\{g_{1}e^{-i\theta }-\omega \gamma -\eta _{1}R(a^{\dagger
}a)-g_{1}T(a^{\dagger }a)\}$ and $\widetilde{\beta}=\{g_{1}e^{-i\theta }-\omega \gamma +\eta _{1}R(a^{\dagger
}a)+g_{1}T(a^{\dagger }a)\}$.

The form of the transformed Hamiltonian $\tilde{H}_{0}$ by considering contributions of the
Rashba and Dresselhaus SOCs is identical with the original Hamiltonian (\ref%
{originalH}) only with Rashba SOC terms. To obtain the solvable Hamiltonian $%
\tilde{H}_{1}$, the transformed Dresselhaus terms $\tilde{H}_{D}$ are
required to be vanished by choosing a proper displacement $\gamma $ and $%
\gamma ^{\ast }$. Within the oscillator basis $|n\rangle $ and the
eigenstates $|\pm z\rangle $ of $\sigma _{z}$, the matrix element $\langle
n,+z|\tilde{H}_{D}|n+1,-z\rangle $ equals to be zero. It yields
\begin{equation}
0=g_{1}e^{i\theta }-\omega \gamma ^{\ast }+\eta _{1}R_{n,n+1}+g_{1}T_{n,n+1}.
\end{equation}%
Since the displacement $\gamma $ ($\gamma ^{\ast }$) is smaller compared
with the unit, it approximately leads to $L_{n}(4\gamma \gamma ^{\ast
})\simeq 1$, $L_{n}^{1}(4\gamma \gamma ^{\ast })\simeq n+1$, and $%
L_{n}^{2}(4\gamma \gamma ^{\ast })\simeq (n+1)(n+2)/2$. One obtains the
simplified equation $g_{1}e^{i\theta }-\omega \gamma ^{\ast }+\gamma ^{\ast
}\Delta +g_{1}e^{-i\theta }=0$, resulting in
\begin{equation}
\gamma \approx \frac{2g_{1}\beta }{(\omega +\Delta )\sqrt{\alpha ^{2}+\beta
^{2}}}.
\end{equation}

We obtain the solvable Hamiltonian $\tilde{H}_{0}$ by considering both of
the Rashba and Dresselhaus SOCs, which retains the Rashba SOC term $a\sigma
_{-}$ and $a^{\dag }\sigma _{+}$. It is so-called GRSOCA.
Different from the RSOC approximation, the effective Rashba SOC strength and
Zeeman energy are renormalized, which leads to
richer physics induced by both types of the SOCs. The effective Hamiltonian
obtained by the variational method is expected to be prior to the original
Hamiltonian $H_{\mathtt{RSOC}}$ (\ref{originalH}) only with the Rashba SOC
terms. The simplicity of the method is based on its analytical eigenstates
and eigenvalues.

One can easily diagonalize the effective Hamiltonian $\tilde{H}_{0}$ in the
basis of $\left\vert n,-z\right\rangle $ \ and $\left\vert
n+1,+z\right\rangle $
\begin{equation}
\tilde{H}_{0}=\left(
\begin{array}{cc}
\omega n+\widetilde{\Delta }_{-,n} & \sqrt{n+1}\widetilde{\alpha}_{n,n+1} \\
\sqrt{n+1}\widetilde{\alpha}_{n,n+1}^{\ast } & \omega (n+1)+\widetilde{\Delta }_{+,n+1}%
\end{array}%
\right) ,  \label{transfham}
\end{equation}%
where the Zeeman energy is transformed into $\widetilde{\Delta }_{\pm
,n}=\eta _{0}\pm f(n)$ with $f(n)=\eta _{1}G_{n,n}+g_{1}F_{n,n}$, and the
effective SOC strength is renormalized as $\widetilde{\alpha}_{n,n+1}=(g_{1}e^{i\theta
}-\omega \gamma ^{\ast })-\eta _{1}R_{n,n+1}-g_{1}T_{n,n+1}$. One obtains
approximately $f(n)\approx \eta _{1}-2g_{1}[e^{i\theta }\gamma ^{\ast
}+n(\gamma e^{-i\theta} +\gamma^{\ast}e^{i\theta })]$, and $\widetilde{\alpha}_{n,n+1}\approx
g_{1}(e^{i\theta}-e^{-i\theta})-\omega\gamma^{\ast}-2\eta _{1}\gamma ^{\ast }$.

Similar to the Hamiltonian $H_{\mathtt{RSOC}}$ in Eq. (\ref{rsoc}) with only
the Rashba SOC, the eigenvalues are obtained as
\begin{eqnarray}  \label{e0}
E_{n,\pm } &=&\omega (n+\frac{1}{2})+\frac{1}{2}[\widetilde{\Delta }_{+,n+1}+\widetilde{\Delta }_{-,n}]
\notag \\
&&\pm \frac{1}{2}\sqrt{[\widetilde{\Delta }_{+,n+1}-\widetilde{\Delta }_{-,n}+\omega ]^{2}+4(n+1)|\widetilde{\alpha}_{n,n+1}|^{2}}.
\notag \\
\end{eqnarray}
And the corresponding eigenstates are expressed in the closed form as
\begin{eqnarray}  \label{varphi+}
\left\vert \varphi _{+,n}\right\rangle =\cos \frac{\theta _{n}}{2}\left\vert
n+1\right\rangle \left\vert +z\right\rangle +\sin \frac{\theta _{n}}{2}%
\left\vert n\right\rangle \left\vert -z\right\rangle ,
\end{eqnarray}

\begin{eqnarray}  \label{varphi-}
\left\vert \varphi _{-,n}\right\rangle =\sin \frac{\theta _{n}}{2}
\left\vert n+1\right\rangle \left\vert +z\right\rangle -\cos \frac{\theta
_{n}}{2}\left\vert n\right\rangle \left\vert -z\right\rangle ,
\end{eqnarray}
where $\theta _{n}=\arccos (\delta _{n}/\sqrt{\delta
_{n}^{2}+4(n+1)|\widetilde{\alpha}_{n,n+1}|^{2}})$ with $\delta _{n}=\omega +\widetilde{\Delta }_{+,n+1}-\widetilde{\Delta }_{-,n}$.

The ground state is $|0,+z\rangle $ with the eigenvalue
\begin{eqnarray}
E_{0}=\eta
_{0}+(\eta_1-2\gamma^*g_1e^{i\theta})e^{-2\gamma\gamma^{*}}.
\end{eqnarray}

As a consequence, the corresponding wave functions of the original Hamiltonian $%
H_{0}$ in Eq.(\ref{originalH}) can be obtained using the unitary
transformation as $\left\vert \Psi _{\pm ,n}\right\rangle =U^{\dagger
}\left\vert \varphi _{\pm ,n}\right\rangle $,
\begin{eqnarray}
\left\vert \Psi _{+,n}\right\rangle &=&\frac{1}{\sqrt{2}}[(\cos \frac{\theta
_{n}}{2}\left\vert -\gamma ,n+1\right\rangle _{d}+\sin \frac{\theta _{n}}{2}%
\left\vert -\gamma ,n\right\rangle _{d})\left\vert +\right\rangle _{x}
\notag \\
&&+(\cos \frac{\theta _{n}}{2}\left\vert \gamma ,n+1\right\rangle _{d}-\sin
\frac{\theta _{n}}{2}\left\vert \gamma ,n\right\rangle _{d})\left\vert
-\right\rangle _{x}],  \notag \\
&&
\end{eqnarray}%
and
\begin{eqnarray}
\left\vert \Psi _{-,n}\right\rangle &=&\frac{1}{\sqrt{2}}[(\sin \frac{\theta
_{n}}{2}\left\vert -\gamma ,n+1\right\rangle _{d}-\cos \frac{\theta _{n}}{2}%
\left\vert -\gamma ,n\right\rangle _{d})\left\vert +\right\rangle _{x}
\notag \\
&&+(\sin \frac{\theta _{n}}{2}\left\vert \gamma ,n+1\right\rangle _{d}+\cos
\frac{\theta _{n}}{2}\left\vert \gamma ,n\right\rangle _{d})\left\vert
-\right\rangle _{x}],  \notag \\
&&
\end{eqnarray}%
where $\left\vert \pm \right\rangle _{x}=(\left\vert +\right\rangle _{z}\pm
\left\vert -\right\rangle _{z})/\sqrt{2}$ is the eigenstate of $\sigma _{x}$%
. Each Laudau state becomes the displaced-Fock state $\left\vert
n\right\rangle $
\begin{equation}
\left\vert \mp \gamma ,n\right\rangle _{d}=e^{\mp (\gamma a^{\dagger
}-a\gamma ^{\ast })}\left\vert n\right\rangle ,
\end{equation}%
which is the displacement transformation of the Fock state $\vert n\rangle$.
Especially it reduces to the coherent state $\left\vert \mp \gamma
,0\right\rangle _{d}=e^{\mp (\gamma a^{\dagger }-a\gamma ^{\ast
})}\left\vert 0\right\rangle $, which can be expanded as a superposition
state of Fock states. Since the Dresselhaus and Rashba SOCs induce infinite $%
n$-th Landau-level states coupling, it is challenge to give eigenstates a
closed form. Fortunately, the novel
displaced-Fock states as a new set of basis states exhibit an improvement over original Fock states.
\begin{figure}[tbp]
\includegraphics[scale=0.53]{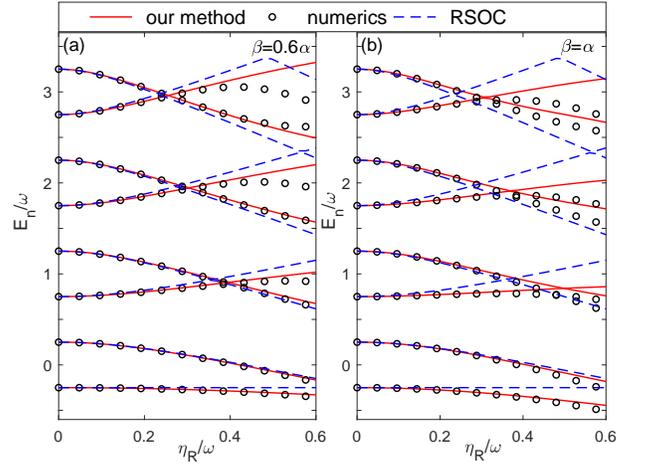}
\caption{ Energy levels $E_{n}/\protect\omega $ obtained analytically (red solid line) as a function of effective
coupling strength $\protect\eta _{R}/\protect\omega =\protect\sqrt{2}\protect%
\alpha /(l_{b}\protect\omega )$ for different ration between the Dresselhaus and Rashba SOCs strength
(a) $\beta /\alpha =0.6$ and (b) $\beta /\alpha =1$.
The results obtained by
the numerical exact diagonalization method (black circles) and under the
RSOC approximation (blue dashed line) are listed for comparison. The
parameters are $\Delta /\protect\omega =0.5$, $l_{b}=1$ and $\omega=1$. }
\label{energy}
\end{figure}

Fig.~\ref{energy} displays the first eight energy levels as a function of
the effective coupling strength $\eta _{R}/\omega=\sqrt{2}\alpha/(l_b\omega)$ for
various values of the Dresselhaus coupling strength $\beta$. In the absence
of the spin-orbit coupling $\eta_R=0$, one observes two separated $n$th
Landau levels induced by the Zeeman energy $\Delta=0.5\omega$, in which the
lower level is the spin-up state and the higher level corresponds to the
spin-down electron state. As $\eta_R$ increases, the higher level of the $n$%
th Landau level state becomes lower due to the hybridization of the $n$th
and $n+1$th displaced-Fock states induced by both types of SOCs. Comparing
with the Rashba SOC approximation, the energy crossing occurs at a larger
value of the coupling strength as a consequence of the Dresselhaus SOC. It
demonstrates that the Dresselhaus SOC enhances Zeeman splitting, while the
Rashba SOC interplay with the Zeeman splitting in opposite ways.

For the ratio between the Dresselhaus and Rashba SOC strengths $%
\beta/\alpha=0.6$, our analytical approach is in good agreement with the
numerical results in a wide range of coupling strength $\eta_R/\omega<0.4$
in Fig.~\ref{energy}(a). There is noticeable deviation of the Rashba SOC
results with the increasing of the coupling strength up to $\eta_R/\omega=0.3
$. When the Dresselhaus and Rashba terms have equal strength with each
other, $\beta=\alpha$, in Fig.~\ref{energy}(b), the deviation becomes more
obvious. Because the Dresselhaus SOC play a more important effect as $%
\eta_R/\omega$ increases, and the Rashba SOC approximation fails. Therefore,
our approach, which takes into account the effects of the Dresselhaus SOC
terms, provides a more accurate analytical expression to the energy spectrum
of the $2$DEG system.

\section{Spin current with a electric field}

Since the competition of the SOC and the Zeeman energy induces an energy
crossing, the spin Hall resonance is closely related to the level crossing.
When an external electric field is applied, the SOC of the $2$DEG induces
the spin Hall effect, which is the transverse spin current response to the
electric field. As the electric field $E$ is applied along the $y$ axis, the
Hamiltonian becomes $H=H_{0}+eEy$ with the original Hamiltonian $H_{0}$
defined in Eq. (\ref{originH0}). Using the replacement of $y $ by $%
y+eE/m\omega ^{2}$ in the oscillator operator $a$, one obtains the quantized
Hamiltonian
\begin{equation}
H=H_{0}+H_{E},H_{E}=-E[\frac{ke}{m\omega }+\frac{e}{\omega }(\alpha \sigma
_{y}+\beta \sigma _{x})],
\end{equation}%
where $H_{0}$ is given in Eq.(\ref{originH0}), and the constant $%
-e^{2}E^{2}/2m\omega ^{2}$ is dropped. Similar to the transformed
Hamiltonian $\tilde{H}_{0}$ in Eq. (\ref{transfham}), we perform the unitary
transformation $U=\exp [\sigma _{x}(a^{\dagger }\gamma -a\gamma ^{\ast })]$
to $H_{E}$, resulting in
\begin{eqnarray}
\tilde{H}_{E}&=&UH_{E}U^{\dag }  \notag \\
&=&-E\frac{ke}{m\omega }-E\frac{\beta e}{\omega }\sigma _{x}-E\frac{\alpha e%
}{\omega }\{\sigma _{y}G(a^{\dagger }a)  \notag \\
&&+i\sigma _{z}[R(a^{\dagger }a)a^{\dagger }-aR(a^{\dagger }a)]\}.
\end{eqnarray}
The wave function for the Hamiltonian with the electric field can be given
to the first-order correction in the perturbation in $\tilde{H}_{E}$ as%
\begin{equation}  \label{wave1}
\left\vert \varphi _{\pm ,n}^{(1)}\right\rangle =\left\vert \varphi _{\pm
,n}\right\rangle +\sum_{n\neq k,l}\frac{\langle \varphi _{l,k}|\tilde{H}_{E}
\left\vert \varphi _{\pm ,n}\right\rangle }{E_{n,\pm }-E_{l,k}}\left\vert
\varphi _{l,k}\right\rangle ,(l=\pm),
\end{equation}
where the eigenvalues $E_{n,\pm }$ and eigenstates $\left\vert \varphi _{\pm
,n}\right\rangle $ are given in Eqs. (\ref{e0})-(\ref{varphi-}).

The charge current operator of a single electron is given by
\begin{eqnarray}
j_{c} &=&-e\upsilon _{x}, \\
\upsilon _{x} &=&\frac{1}{i}[x,H]=\frac{p_x}{m}+\omega y+\alpha \sigma
_{y}+\beta \sigma _{x},
\end{eqnarray}%
and the spin-$z$ component current operator is
\begin{eqnarray}
j_{s}^{z} &=&\frac{\hbar }{2}(S^{z}\upsilon _{x}+\upsilon _{x}S^{z})  \notag
\\
&=&\frac{1 }{2}[\sqrt{\frac{\omega }{2m}}(a^{\dag }+a)-\frac{eE}{%
m\omega }]\sigma _{z}.
\end{eqnarray}
The average current density of the $N_{e}$ electron system is given by
\begin{equation}
I_{c(s)}=\frac{1}{L_{x}L_{y}}\sum_{nl}\langle j_{c(s)}\rangle
_{nl}f(E_{nl}),(l=\pm 1),
\end{equation}%
where $f(E_{nl})$ is the Fermi distribution function, and $%
N_{e}=\sum_{nl}f(E_{nl})$. The charge Hall conductance is
\begin{equation}
G_{c(s)}=I_{c(s)}/E.
\end{equation}

\begin{figure}[tbp]
\includegraphics[trim=120 10 130 10,width=0.55\linewidth]{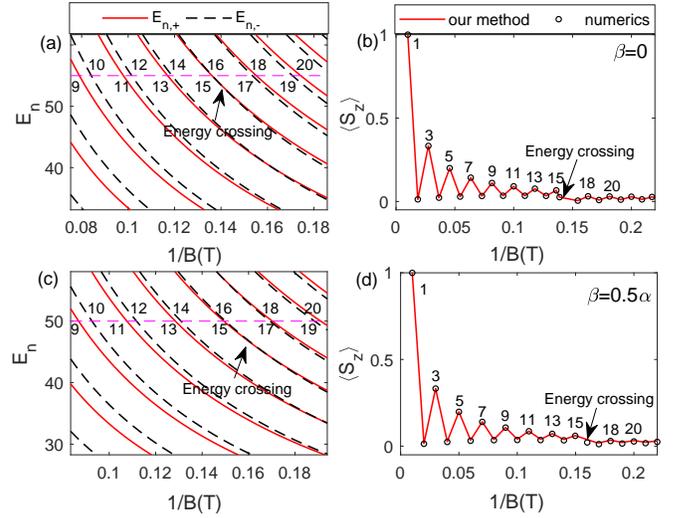}
\caption{ Energy levels $E_n$ and average spin $\langle S_z\rangle$ obtained
analytically for an electron as a function of $1/B$ for $\protect\beta=0$%
(a)(b) and $\protect\beta/\protect\alpha=0.5$ (c) (d) with $\Delta/\protect%
\omega=0.5$ and $\alpha=0.25\omega$. The results of $\langle\protect\sigma_z\rangle$ obtained by the numerical
exact diagonalization (black circles) are listed for comparison. }
\label{spin current}
\end{figure}
Under the first-order perturbation, the corresponding spin/charge current
can be expressed as $\langle j_{c(s)}\rangle _{\pm ,n}=\langle
j_{c(s)}^{(0)}\rangle _{\pm n}+\langle j_{c(s)}^{(1)}\rangle _{\pm n}$,
where
\begin{eqnarray}
\langle j_{c(s)}^{(0)}\rangle _{\pm n} =\left\langle \varphi
_{\pm,n}\right\vert Uj_{c(s)}U^{\dag }\left\vert \varphi
_{\pm,n}\right\rangle ,
\end{eqnarray}
\begin{eqnarray}\label{js1}
\langle j_{c(s)}^{(1)}\rangle _{\pm n} &=&\sum\limits_{n\neq k,l}\frac{%
\left\langle \varphi _{\pm ,n}\right\vert \tilde{H}_{E}\left\vert \varphi
_{l,k}\right\rangle \left\langle \varphi _{l,k}\right\vert Uj_{c(s)}U^{\dag
}\left\vert \varphi _{\pm ,n}\right\rangle }{E_{k,l}-E_{n,\pm }}  \notag \\
&&+h.c.
\end{eqnarray}%
Under the zeroth approximation, one obtains analytical solutions
\begin{eqnarray}
\left\langle j_{c}^{(0)}\right\rangle _{\pm n}=\frac{e^{2}E}{hN_{\phi }}%
L_{x}L_{y}, \langle j_{s}^{z(0)}\rangle _{\pm ,n}=-\frac{eE}{2m\omega }%
\left\langle \sigma _{z}\right\rangle _{\pm n},
\end{eqnarray}
where $\left\langle \sigma _{z}\right\rangle _{\pm n}$ is given in the
Appendix B. With the average current density $I_{s}^{z}$, the spin Hall
conductance can be derived under the zeroth order correction by
\begin{equation}
G_{s}^{z(0)}=-\frac{\langle S_{z}\rangle}{E}\frac{eE}{m\omega }=-\frac{%
\langle S_{z}\rangle G_{c}}{e},
\end{equation}%
\begin{equation}
\langle S_{z}\rangle=\sum_{nl}\frac{1}{2}\left\langle \sigma
_{z}\right\rangle _{nl}f(E_{n,l}),(l=\pm ).
\end{equation}%
And the Hall conductance is given as in the Appendix B
\begin{equation}\label{chargeGc}
G_{c}=e^{2}N_{e}/(2\pi N_{\phi }),
\end{equation}
which is only dependent on the filling factor $N_{e}/N_{\phi }$ with $N_{\phi }=L_{x}L_{y}eB/(hc)$ .

Fig.~\ref{spin current} shows energy levels and the spin polarization $%
\langle S_z\rangle$ under the zeroth approximation. It is observed that the
energy $E_{n,+}$ with spin-up state firstly enters into the Fermi energy
region, then it gives rise to the energy $E_{n+1,-}$ with spin-down state in
Fig.~\ref{spin current}(a)(c). As the energy gap between $E_{n,+}$ and $%
E_{n+1,-}$ becomes smaller, it yields energy crossing at certain magnetic
field $B_0$, which is given by $E_{n+1,-}=E_{n,+}$ in Eq.(\ref{e0}). When
the magnetic field exceeds the critical value $B_0$, the spin-down state
with $E_{n,-}$ emerge firstly, and then the spin-up state with $E_{n+1,+}$
enters into the Fermi energy region. The corresponding expected value of $%
\langle S_{z}\rangle $ is calculated in Fig.~\ref{spin current}(b)(d). It
reaches maxima at odd integers $n$, and minima at even integers $n $. A
discontinuous jump occurs at $B_{0}$. Below the critical value $B_{0}$, the
maximal value of $\langle \sigma _{z}\rangle $ occurs at even integers $n $.
The jump of the spin polarization ascribes to the energy crossing of two
eigenstates with almost opposite spins. Especially, when only Rashba SOC is
considered ($\beta=0$), one obtains the constraint condition for the energy crossing
\begin{equation}
2\omega =\sqrt{(\omega-\Delta )^{2}+4(n+1)\eta_R^{2}} +\sqrt{(\omega-\Delta )^{2}+4(n+2)\eta_R^{2}},
\end{equation}
with the displacement $\gamma =0$ (see the Appendix
C). It recovers results with only the Rashba coupling~\cite{shen2004}. By
comparing to the results with only Rashba SOC, the critical value of $1/B_0$
shifts to a larger value in Fig.~\ref{spin current}(d). It demonstrates that
the Dresselhaus SOC plays a vital role in suppressing the energy crossing,
which is different from the effects of the Rashba SOC.

\begin{figure}[tbp]
\includegraphics[width=8.5cm,height=8cm]{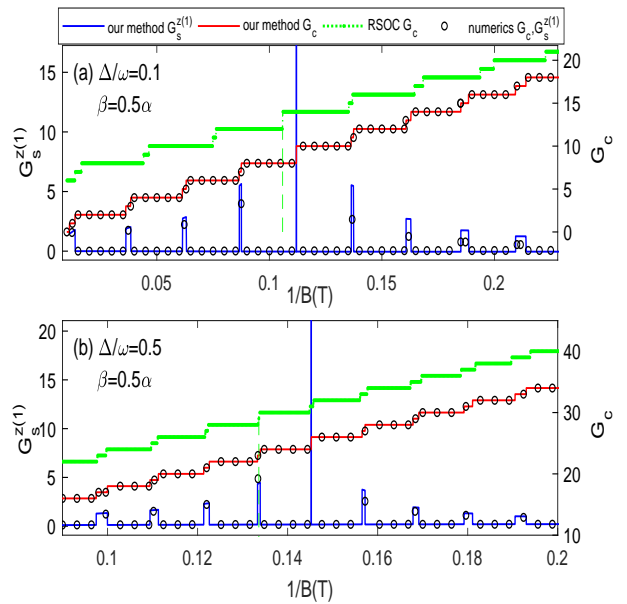}
\caption{Charge conductance $G_{c}$ and spin Hall conductance $G_{s}^{z(1)}$ obtained
with first-order corrections as a function of $1/B$ for
different Zeeman splitting energy (a) $\Delta/\protect\omega=0.1$ and (b) $%
\Delta/\protect\omega=0.5$. The ratio between the Dresselhaus and Rashba SOCs is
$\beta/\protect\alpha=0.5$ with the Rashba SOC strength $\alpha=0.25\omega$. The results obtained by
the numerical exact diagonalization (black circles) and under the RSOC
approximation (blue dotted line) are listed for comparison. The external
electronic field is $E/\protect\omega=0.1N/C$. }
\label{spin conductance}
\end{figure}

In presence of the electric field, the spin Hall conductance of the spin-$z$
component current is the most interesting. Fig.~\ref{spin conductance} shows
the charge conductance $G_{c}$ in Eq. (\ref{chargeGc}) and the spin Hall conductance $G_{s}^{z(1)}$
obtained by the first-order corrections in Eq. (\ref{js1}). A series of plateaus in the charge $%
G_{c}$ are visible, and a jump between two plateaus is observed at the
critical magnetic field, where the spin conductance $G_{s}^{z(1)} $ becomes
divergent with a resonant peak. The resonance ascribes to the interference
of two degenerate levels near the Fermi energy. The resonance point
coincides with the jump point of $\langle S_z\rangle$ with the energy
crossing. By comparing to the behaviors with only the Rashba SOC, the charge
$G_{c}$ and spin Hall conductance $G_{s}^{z(1)}$ exhibit a shift value of
the resonant point, which is induced by the Dresselhaus SOC effects. For a
large Zeeman splitting energy $\Delta/\omega=0.5$, the resonant point shifts
to a larger value of $1/B_0$ in Fig.~\ref{spin conductance}(b). It
demonstrate that the SOC interactions and the Zeeman splitting play an
opposite role in the energy-levels crossing. Fortunately, the charge and
spin conductance obtained by first-order approximation agree well with the
numerical results, exhibiting the validity of our approach.

\section{Conclusion}

When both the Rashba and Dresselhaus spin-orbit couplings are considered, we
find the single electron Hamiltonian in two-dimensional system subjected to
a perpendicular magnetic field can map onto an anisotropic Rabi model. We
perform the generalized Rashba SOC approximation using the displacement unitary transformation,
and keep the single Landau
level (nearest neighbor Landau level mixing) matrix element for even (odd)
coupling function, and a solvable Hamiltonian is obtained in a similar form
as that with only the Rashba term. The strengths of the both types of SOCs
and Zeeman splitting are absorbed in the displacement-shift variable. With
comparing the numerical diagonalization, our method provides accurate energy
levels up to a large Zeeman splitting. As a consequence of the Dresselhaus
and Rashba SOCs, each Landau state becomes a displaced-Fock state, which has
a displacement shift by comparing to the original Harmonic oscillator Fock
state. With the analytical solved eigenstates, the spin current displays a
jump at a larger value of the inverse of the magnetic field, which
demonstrates that the Dresselhaus SOC plays an opposite way in the energy
splitting by comparing to the Rashba SOC. Moreover, in the presence of an
electric field, the spin Hall conductance obtained by the first-order
corrections diverges at the resonant point, and a series of plateaus of the
charge conductance are observed, which fit well with numerical results. In
conclusion, our method provides an easy-to-implement analytical solution to
the 2DEGs with considering all SOCs in which all the coupling strengths,
including Rashba, Dresselhaus, and Zeeman splitting, are described by the
displacement shift. This solution could be potentially useful in the future
studies of the quantum version of the spin Hall effects and the interacting
fractional quantum Hall systems.

\acknowledgments
This work was supported by National Natural Science Foundation of China
(Grants No.12075040, No.11875231, and No.11974064), and by the Chongqing
Research Program of Basic Research and Frontier Technology (Grants
No.cstc2020jcyj-msxmX0890).

\appendix

\section{Deviation of the Hamiltonian by the displacement
transformation}

We perform the unitrary transformation $U=\exp [\sigma _{x}(a^{\dagger
}\gamma -a\gamma ^{\ast })]$ to the Hamiltonian $H_{0}$ in Eq. (\ref{originalH}). One
easily obtains $UaU^{\dagger }=a-\gamma \sigma _{x}$ and $Ua^{\dagger
}U^{\dagger }=a^{\dagger }-\gamma ^{\ast }\sigma _{x}$. The first and second
terms of $H_{0}$ in Eq. (\ref{originalH}) can be transformed into
\begin{equation}
Ua^{\dagger }aU^{\dagger }=a^{\dagger }a-\sigma _{x}(a^{\dagger }\gamma
+a\gamma ^{\ast })+\gamma \gamma ^{\ast },
\end{equation}%
and
\begin{eqnarray}
&&U\sigma _{z}U^{\dagger } =\sigma _{z}\{1+\frac{1}{2}\sigma
_{z}[2(a^{\dagger }\gamma -a\gamma ^{\ast })]^{2}+...\}  \notag \\
&&-i\sigma _{y}\{2(a^{\dagger }\gamma -a\gamma ^{\ast })+\frac{1}{3!}%
[2(a^{\dagger }\gamma -a\gamma ^{\ast })]^{3}+...\}  \notag \\
&=&\sigma _{z}\cosh [2(a^{\dagger }\gamma -a\gamma ^{\ast })]-i\sigma
_{y}\sinh [2(a^{\dagger }\gamma -a\gamma ^{\ast })].\notag \\
\end{eqnarray}
Meanwhile, two SOCs terms of $H_{0}$ are derived explicitly as
\begin{equation}
U\sigma _{x}(a^{\dagger }e^{-i\theta }+ae^{i\theta })U^{\dagger }=\sigma
_{x}(a^{\dagger }e^{-i\theta }+ae^{i\theta })-(\gamma ^{\ast }e^{-i\theta
}+\gamma e^{i\theta }),
\end{equation}%
and%
\begin{eqnarray*}
&&Ui\sigma _{y}(a^{\dagger }e^{i\theta }-ae^{-i\theta })U^{\dagger }\\
&&=i\sigma _{y}B-\sigma _{z}[2AB-(\gamma e^{-i\theta }-\gamma ^{\ast
}e^{i\theta })] \\
&&+\frac{1}{2!}i\sigma _{y}[4A^{2}B-4A(\gamma e^{-i\theta }-\gamma ^{\ast
}e^{i\theta })] \\
&&-\frac{1}{3!}\sigma _{z}[8A^{3}B-12A^{2}(\gamma e^{-i\theta }-\gamma
^{\ast }e^{i\theta })^{2}] \\
&&+\frac{1}{4!}i\sigma _{y}[16A^{4}B-32A^{3}(\gamma e^{-i\theta }-\gamma
^{\ast }e^{i\theta })]+... \\
&=&i\sigma _{y}[\cosh (2A)B-(\gamma e^{-i\theta }-\gamma ^{\ast }e^{i\theta
})\sinh (2A)] \\
&&-\sigma _{z}[\sinh (2A)B-(\gamma e^{-i\theta }-\gamma ^{\ast }e^{i\theta
})\cosh (2A)]
\end{eqnarray*}%
where the operators are given by $A=a^{\dagger }\gamma -a\gamma ^{\ast }$
and $B=a^{\dagger }e^{i\theta }-ae^{-i\theta }$. Thus, the transformed
Hamiltonian is given in terms of $H_{0}^{\prime }$ and $H_{1}^{\prime }$ in Eqs. (\ref{transh0}) and (\ref{transh1}).

By expanding the even and odd functions $\cosh [2(a^{\dagger }\gamma
-a\gamma ^{\ast })]$ and $\sinh [2(a^{\dagger }\gamma -a\gamma ^{\ast })]$,
the corresponding coefficients are derived as%
\begin{eqnarray}
G_{n,n} &=&\left\langle n\right\vert \cosh [2(a^{\dagger }\gamma -a\gamma
^{\ast })]\left\vert n\right\rangle   \notag \\
&=&\frac{1}{2}\left\langle n\right\vert \{\exp [2(a^{\dagger }\gamma
-a\gamma ^{\ast })]+\exp [-2(a^{\dagger }\gamma -a\gamma ^{\ast
})]\}\left\vert n\right\rangle   \notag \\
&=&e^{-2\gamma \gamma ^{\ast }}\sum_{i=0}^{n}(-1)^{n-i}\frac{n!}{%
(n-i)!(n-i!)i!}(4\gamma \gamma ^{\ast })^{n-i}  \notag \\
&=&e^{-2\gamma \gamma ^{\ast }}L_{n}(4\gamma \gamma ^{\ast }),
\end{eqnarray}%
and
\begin{eqnarray}
&&\sqrt{n+1}R_{n+1,n}=\left\langle n+1\right\vert \sinh [2(a^{\dagger
}\gamma -a\gamma ^{\ast })]\left\vert n\right\rangle   \notag \\
&=&\frac{1}{2}\left\langle n\right\vert \{\exp [2(a^{\dagger }\gamma
-a\gamma ^{\ast })]-\exp [-2(a^{\dagger }\gamma -a\gamma ^{\ast
})]\}\left\vert n\right\rangle   \notag \\
&=&\frac{2\gamma e^{-2\gamma \gamma ^{\ast
}} }{\sqrt{n!(n+1)!}}\sum_{i=0}^{n}(-1)^{n-i}\frac{n!(n+1)!}{(n-i)!(n+1-i!)i!}(4\gamma \gamma
^{\ast })^{n-i}  \notag \\
&=&\frac{2\gamma }{\sqrt{n+1}}e^{-2\gamma \gamma ^{\ast }}L_{n}^{1}(4\gamma
\gamma ^{\ast }).
\end{eqnarray}

\section{Spin current under the zeroth corrections}

In presence of the electric field, the spin current is derived as
\begin{eqnarray}
\left\langle j_{s}^{z}\right\rangle _{\pm n} =\frac{1 }{2}\langle
\varphi _{\pm ,n}|U[\sqrt{\frac{\hbar \omega }{2m}}(a^{\dag }+a)-\frac{eE}{%
m\omega}]\sigma _{z}U^{\dagger }\left\vert \varphi _{\pm
,n}\right\rangle.\notag \\
\end{eqnarray}
Using the eigenstates $\left\vert \varphi _{\pm
,n}\right\rangle $ in Eqs. (\ref{varphi+}) and (\ref{varphi-}), one obtains
\begin{eqnarray}
&&\langle (a^{\dag }+a)\sigma _{z}\rangle _{+n}=\left\langle \varphi
_{+,n}\right\vert U(a^{\dag }+a)\sigma _{z}U^{\dag }\left\vert \varphi
_{+,n}\right\rangle  \notag \\
&=&\left\langle \varphi _{+,n}\right\vert [(a^{\dag }+a)-(\gamma +\gamma
^{\ast })\sigma _{x}]U\sigma _{z}U^{\dagger }\left\vert \varphi
_{+,n}\right\rangle  \notag \\
&=&\left\langle n\right\vert (a^{\dagger }+a)\sinh [2(a^{\dagger }\gamma
-a\gamma ^{\ast })]\sin ^{\ast }\frac{\theta _{n}}{2}\cos \frac{\theta _{n}}{%
2}\left\vert n+1\right\rangle  \notag \\
&&-\left\langle n+1\right\vert (a^{\dagger }+a)\sinh [2(a^{\dagger }\gamma
-a\gamma ^{\ast })]\cos ^{\ast }\frac{\theta _{n}}{2}\sin \frac{\theta _{n}}{%
2}\left\vert n\right\rangle  \notag \\
&=&0,
\end{eqnarray}
and
\begin{equation}
\langle (a^{\dag }+a)\sigma _{z}\rangle _{-n}=\left\langle \varphi
_{-,n}\right\vert U(a^{\dag }+a)\sigma _{z}U^{\dag }\left\vert \varphi
_{-,n}\right\rangle =0.
\end{equation}
So the spin current is simplified as
\begin{eqnarray}
\left\langle j_{s}^{z}\right\rangle _{\pm n}=-\frac{eE}{2m\omega }\left\langle \sigma _{z}\right\rangle _{\pm n}.
\end{eqnarray}
where the average value of $\left\langle \sigma _{z}\right\rangle _{\pm n}$ are derived in the following
\begin{eqnarray}
&&\left\langle \sigma _{z}\right\rangle _{+n}=\langle \varphi _{+,n}|U\sigma
_{z}U^{\dagger }\left\vert \varphi _{+,n}\right\rangle  \notag \\
&&=\cos \frac{\theta _{n}}{2}\cos ^{\ast }\frac{\theta _{n}}{2}%
G_{n+1,n+1}-\sin \frac{\theta _{n}}{2}\sin ^{\ast }\frac{\theta _{n}}{2}%
G_{n,n}  \notag \\
&&-\sqrt{n+1}(\sin \frac{\theta _{n}}{2}\cos ^{\ast }\frac{\theta _{n}}{2}%
R_{n+1,n}+\sin ^{^{\ast }}\frac{\theta _{n}}{2}\cos \frac{\theta _{n}}{2}%
R_{n,n+1}),  \notag \\
&&
\end{eqnarray}
and
\begin{eqnarray}
&&\left\langle \sigma _{z}\right\rangle _{-n}=\langle \varphi _{-,n}|U\sigma
_{z}U^{\dagger }\left\vert \varphi _{-,n}\right\rangle  \notag \\
&&=\sin \frac{\theta _{n}}{2}\sin ^{\ast }\frac{\theta _{n}}{2}%
G_{n+1,n+1}-\cos \frac{\theta _{n}}{2}\cos ^{\ast }\frac{\theta _{n}}{2}%
G_{n,n}  \notag \\
&&+\sqrt{n+1}(\cos \frac{\theta _{n}}{2}\sin ^{\ast }\frac{\theta _{n}}{2}%
R_{n+1,n}+\cos ^{\ast }\frac{\theta _{n}}{2}\sin \frac{\theta _{n}}{2}%
R_{n,n+1}).  \notag \\
&&
\end{eqnarray}

Meanwhile, the charge current $j_{c}=-ev_{x}$ can be expressed in terms of the harmonic
oscillator $a$ ($a^{\dagger }$) as
\begin{equation}
j_{c}=-e[\sqrt{\frac{\hbar \omega }{2m}}(a^{\dagger }+a)+\alpha \sigma
_{y}+\beta \sigma _{x}-\frac{eE}{m\omega }].
\end{equation}%
The average value of the charge current is
given by%
\begin{eqnarray}
\left\langle j_{c}\right\rangle _{\pm n} &=&\langle \varphi _{\pm
,n}|Uj_{c}U^{\dag }\left\vert \varphi _{\pm ,n}\right\rangle  \notag \\
&=&\left\langle \varphi _{\pm ,n}\right\vert \frac{e^{2}E}{m\omega }%
\left\vert \varphi _{\pm ,n}\right\rangle =\frac{e^{2}E}{2\pi N_{\phi }}%
L_{x}L_{y},
\end{eqnarray}%
where $Uj_{c}U^{\dag }=-e[\sqrt{\frac{\omega }{2m}}(a^{\dagger }+a)-\sqrt{%
\frac{\omega }{2m}}(\gamma +\gamma ^{\ast })\sigma _{x}+\alpha \sigma
_{y}\cosh (2A)+\alpha i\sigma _{z}\sinh(2A)+\beta \sigma _{x}-eE/(m\omega)]$.
Then we obtain the charge Hall conductance $G_{c}=I_{c}/E$ as
\begin{equation}
G_{c}=\frac{e^{2}E}{2\pi EN_{\phi }}\sum_{nl}f(E_{nl})=\frac{e^{2}N_{e}}{2\pi N_{\phi }}.
\end{equation}

\section{energy-crossing conditions}

Since the $n+1$-th Laudau level of spin down and the $n$-th Laudau level of spin-up
becomes crossing near the Fermi energy, resulting in the resonance peak of the spin Hall conductance
at the energy crossing point. It leads to the energy crossing at certain magnetic
field $B_0$, which satisfies $E_{n+1,-}=E_{n,+}$. It yields
\begin{eqnarray}
&&2\omega +f(n+2)-2f(n+1)+f(n)  \notag \\
&&=\sqrt{[f(n+1)+f(n)+\omega ]^{2}+4(n+1)|\widetilde{\alpha}_{n,n+1}|^{2 }}  \notag \\
&&+\sqrt{[f(n+2)+f(n+1)+\omega ]^{2}+4(n+2)|\widetilde{\alpha}_{n+1,n+2}|^{2 }}.  \notag \\
\end{eqnarray}%
Especially, when the Dresshaul SOC is neglected by setting $\beta =0$, the
displacement variable reduces into $\gamma =0$. One can simplify the parameters as $%
R_{n,n+1}=R_{n+1,n}=0$, $T_{n,n+1}=-i\alpha/\sqrt{\alpha^2+\beta^2} $, $\eta _{0}=\omega /2$, $\eta
_{1}=-\Delta /2$, $f(n)=-\Delta /2$, and $\widetilde{\alpha}_{n,n+1}=i\eta _{R\text{ }}$.
It leads to the eigenvalues
\begin{equation}
E_{n,+}=\omega n+\omega+\frac{1}{2}\sqrt{(\omega-\Delta)^{2}+4(n+1)\eta _{R}^{2}},
\end{equation}%
\begin{equation}
E_{n+1,-}=\omega (n+1)+\omega-\frac{1}{2}\sqrt{%
(\omega-\Delta )^{2}+4(n+2)\eta _{R}^{2}}.
\end{equation}%
Thus, the energy-levels crossing is given by $E_{n+1,-}-E_{n,+
}=0 $, resulting in
\begin{equation}
2\omega =\sqrt{(\omega-\Delta)^{2}+4(n+1)\eta _{R}^{2}}+\sqrt{(\omega-\Delta)^{2}+4(n+2)\eta _{R}^{2}}.
\end{equation}

\end{document}